# Scalar-Tensor Cosmological Models

A. Serna and J. M. Alimi

*Laboratoire d'Astrophysique Extragalactique et de Cosmologie,*

*CNRS, URA 173, Observatoire de Paris-Meudon, 92195-Meudon, France*

(October 31, 1995)

## Abstract

We analyze the qualitative behaviors of scalar-tensor cosmologies with an arbitrary monotonic $\omega(\Phi)$ function. In particular, we are interested on scalar-tensor theories distinguishable at early epochs from General Relativity (GR) but leading to predictions compatible with solar-system experiments. After extending the method developed by Lorentz-Petzold and Barrow, we establish the conditions required for convergence towards GR at $t \to \infty$. Then, we obtain all the asymptotic analytical solutions at early times which are possible in the framework of these theories. The subsequent qualitative evolution, from these asymptotic solutions until their later convergence towards GR, has been then analyzed by means of numerical computations. From this analysis, we have been able to establish a classification of the different qualitative behaviors of scalar-tensor cosmological models with an arbitrary monotonic $\omega(\Phi)$ function.

PACS number(s): 04.50.+h, 04.80.+z, 98.80.Cq, 98.80.Hw

Typeset using REVTEX

# I. INTRODUCTION

The Einstein Equivalence Principle generates a whole class of theories (called metric theories) describing gravitational interaction. The distinction among these theories lies in the number and type (tensorial, vectorial, scalar) of additional gravitational fields, other than the metric tensor $g_{\mu\nu}$, they contain. From a theoretical point of view, the most natural alternatives to General Relativity (GR) are scalar-tensor theories, which contain an additional scalar field, $\phi$, the relative importance of which is determined by an arbitrary coupling function $\omega(\phi)$ [1–3]. In recent years, this class of metric theories has received a renewed interest specially in cosmology [4–11], because it provides a natural (non-fine-tuned) way to restore the original ideas of inflation while avoiding the cosmological difficulties coming from the vacuum-dominated exponential expansion obtained in GR. Scalar-tensor theories also arise in current theoretical attempts at deepening the connection between gravitation and the other interactions. For example, in modern revivals of the Kaluza-Klein theory and in supersymmetric theories with extra dimensions, one or several scalar fields arise in the compactification of these extra dimensions [12–18]. Furthermore, scalar-tensor theories may also appear as a low-energy limit of superstring theories (see, e.g., Ref. [19]).

Observational constraints upon scalar-tensor theories can be derived from their effects on solar-system experiments [20]. These quasi-stationary weak field tests imply that any alternative gravity theory must be, in that regime, very close to GR. However, these tests say nothing about how correct is a theory when gravitational forces are very strong. In general, the physical conditions in the early universe and the subsequent cosmological evolution in scalar-tensor theories could be very different from that obtained in the framework of GR. Since a wide class of these theories exhibits an attractive mechanism toward Einstein's theory [8,9,21–23], they could be compatible with present solar-system experiments in spite of their very different predictions at early times. *The full consequences of an arbitrary function $\omega(\phi)$ will be only known when the cosmological solutions of the field equations are studied* [3]. Analytical or numerical solutions for the scalar-tensor cosmological models are well known in the framework of some particular theories proposed in the literature. This is the case of *Brans-Dicke's* theory [24–29], *Barker's* constant-$G$ theory [30,31], *Bekenstein's* variable rest mass theory [32–34], or *Schmidt-Greiner-Heinz-Muller's* theory [35,36]. On the contrary, the properties of more general scalar-tensor theories still remain poorly known. Some important progress have been nevertheless made in the last years. For example, Burd and Coley [37] have used a dynamical system treatment to analyze the qualitative behavior of those models which result when a constant (Brans-Dicke) coupling function is perturbed by a slight dependence on the scalar field (e.g., $\omega(\phi) = \omega_0 + \phi^{-\alpha}$, with $\phi \to \infty$ and $\alpha > 0$).

A different approach has raised more recently from the work by Barrow [28]. In that work, he improved the method by Lorentz-Petzold [38] to solve, in an easy way, the cosmological equations of any scalar-tensor theory with a specified form of $\omega(\phi)$. By using this method, it has been possible to analyze the behavior at early times of some families of theories [28,39,40]. However, an exhaustive study of the properties of any possible scalar-tensor cosmology still is not available in the literature. The aim of this paper is to perform such a study in the hope of finding models with early behaviors qualitatively different from those found in the previous particular cases. To that end, we will restrict ourselves to theories with a vanishing cosmological constant term, and where $\omega(\phi)$ is a monotonic, but arbitrary, function of $\phi$.

The paper is arranged as follows. We begin outlining the Scalar-Tensor theories and we then show how to build up homogeneous and isotropic cosmological models in their framework (Sect. II). After establishing the conditions required for the convergence towards GR (Sect. III) we will present all the possible asymptotic solutions at early times which can be found in this class of theories (Sect. IV). The radiation-dominated evolution from these early behaviors until their later convergence towards GR will be obtained (Sect. V) for a wide class of scalar-tensor theories. Finally, conclusions and a summary of our results are given in Sec. VI.

## II. SCALAR-TENSOR GRAVITY THEORIES

### A. Field equations: the Jordan and Einstein frames

The most general action describing a massless scalar-tensor theory of gravitation is [1–3]

$$S = \frac{1}{16\pi} \int \left( \phi \mathcal{R} - \frac{\omega(\phi)}{\phi} \phi_{,\mu} \phi^{,\mu} \right) \sqrt{g} d^4x + S_M \quad (1)$$

where $\mathcal{R}$ is the curvature scalar of the metrics $g_{\mu\nu}$, $g \equiv det(g_{\mu\nu})$, $\phi$ is the scalar field, and $\omega(\phi)$ is an arbitrary coupling function determining the relative importance of the scalar field. The action (1) has been expressed in terms of the metric tensor $g_{\mu\nu}$, called the "Jordan frame" [41,42], to which matter is universally coupled. In this frame, $g_{\mu\nu}$ is measured by using non-gravitational rods and clocks.

The dynamics are sometimes more simply described by using the "Einstein frame" [41,42,23]

$$S_* = \frac{c^4}{16\pi G_*} \int (\mathcal{R}_* - 2\varphi_{,\mu} \varphi^{,\mu}) \sqrt{g_*} \frac{d^4x}{c} + S_M^* \quad (2)$$

where Newton's constant $G_*$ and the metrics $g_{\mu\nu}^*$ are now measured by using purely gravitational clocks. This frame is obtained from (1) by a conformal transformation

$$g_{\mu\nu} = A^2(\varphi) g_{\mu\nu}^*$$
$$A^2(\varphi) = (G_* \phi)^{-1} \quad (3)$$

where $A(\varphi)$ is an arbitrary function related to $\omega(\phi)$ by

$$\alpha = |3 + 2\omega|^{-1/2} = \frac{\partial \ln A}{\partial \varphi} \quad (4)$$

Since our measures are based on non-purely gravitational rods and clocks, observable quantities are those written in the Jordan frame. Comparison between theory and observations must be then performed by using this physical frame, which will be that used throughout this paper.

The variation of Eq. (1) with respect to $g_{\mu\nu}$ and $\phi$ leads to the field equations:

$$\mathcal{R}_{\mu\nu} - \frac{1}{2}g_{\mu\nu}\mathcal{R} = -\frac{8\pi}{\phi}T_{\mu\nu} - \frac{\omega}{\phi^2}\left(\phi_{,\mu}\phi_{,\nu} - \frac{1}{2}g_{\mu\nu}\phi_{,\alpha}\phi^{,\alpha}\right)$$
$$-\frac{1}{\phi}(\phi_{,\mu;\nu} - g_{\mu\nu}\Box\phi) \qquad (5a)$$

$$(3+2\omega)\Box\phi = 8\pi T - \omega'\phi_{,\alpha}\phi^{,\alpha} \qquad (5b)$$

which satisfy the usual conservation law

$$T^{\mu\nu}_{;\nu} = 0 \qquad (6)$$

where $T^{\mu\nu}$ is the energy-momentum tensor, $\omega'$ denotes $d\omega/d\phi$ and $\Box\phi \equiv g^{\mu\nu}\phi_{,\mu;\nu}$.

### B. Cosmological models

In order to build up cosmological models, we consider a homogeneous and isotropic universe. The line-element has then a Robertson-Walker form:

$$ds^2 = -c^2 dt^2 + R^2(t)\left[\frac{dr^2}{1-Kr^2} + r^2 d\Omega^2\right] \qquad (7)$$

and the energy-momentum tensor corresponds to that of a perfect fluid

$$T^{\mu\nu} = (\rho + P/c^2)u_\mu u_\nu + Pg_{\mu\nu} \qquad (8)$$

where $K = 0, \pm 1$, $R(t)$ is the scale factor, $\rho$ and $P$ are the energy-mass density and pressure, respectively, and $u_\mu$ is the 4-velocity of the fluid. The field equations (5a) and (5b) then become

$$\frac{8\pi}{3\phi}\rho = \frac{c^2 k}{R_0^2 a^2} + \frac{\dot{a}^2}{a^2} - \frac{\omega}{6}\frac{\dot{\phi}^2}{\phi^2} + \frac{\dot{a}\dot{\phi}}{a\phi}$$

$$\ddot{\phi} + 3\frac{\dot{a}}{a}\dot{\phi} = \frac{1}{(3+2\omega)}[8\pi(\rho - 3P/c^2) - \omega'\dot{\phi}^2] \qquad (9)$$

where $a \equiv R/R_0$ is the dimensionless scale factor and dots mean time derivatives. In addition, we have the conservation equation

$$d(\rho a^3) + (P/c^2)da^3 = 0 \qquad (10)$$

Exact solutions of Eqs. (9) can be found, whatever the sign of $3+2\omega$ is, by extending the method developed by Lorentz-Petzold [38] and Barrow [28]. During the vacuum ($\rho = P = 0$) and radiation-dominated epochs, the energy-mass density is given by $\rho = 3P/c^2 \propto a^{-4}$. By introducing a conformal time, $\tau$, and the Lorentz-Petzold [38] variable, $y$, defined as

$$ad\tau = dt \tag{11}$$

$$y = a^2 \phi \tag{12}$$

equations (9) then transform to

$$y'^2 = 4\Gamma y + \tfrac{1}{3}\phi'^2 a^4 (3 + 2\omega) \tag{13}$$

$$\phi'' + 2\tfrac{a'}{a}\phi' = -\tfrac{\phi'^2 (d\omega/d\phi)}{3+2\omega} \tag{14}$$

where primes denote $d/d\tau$ and $\Gamma \equiv 8\pi\rho_{rad}/3$ ($\rho_{rad}$ is the energy-mass density at the end of the radiation-dominated era). As usual, we have considered that $k/a^2$ can be neglected in the radiation-dominated field equations and, hence, that zero curvature models provide a good description of the early evolution of the universe.

Integration of Eq. (14) gives

$$\phi' a^2 = \text{sign}(3+2\omega) 2\sqrt{3}\Gamma A \,|3+2\omega|^{-1/2} \tag{15}$$

where $A$ is a constant and where we have allowed for negative values of $(3+2\omega)$.

By using Eq. (15), the integration of Eqs. (13) and (14) gives

$$y = \Gamma[(\tau + \tau_0)^2 - \text{sign}(3+2\omega) A^2] \tag{16}$$

and

$$\int |3+2\omega|^{1/2} \frac{d\phi}{\phi} = \begin{cases} \sqrt{3}\ln(f) & (3+2\omega > 0) \\ 2\sqrt{3}\{\arctan[(\tau+\tau_0)/A] - \pi/2\} & (3+2\omega < 0) \end{cases} \tag{17}$$

where

$$f = \frac{\tau + \tau_0 - A}{\tau + \tau_0 + A} \qquad (18)$$

$\tau_0$ being an integration constant.

Obviously, for a given form of $\omega(\phi)$, Eq. (17) can be integrated to obtain $\phi(\tau)$. Equations (12) and (16) give then $a(\tau)$, and the relation $\tau(t)$ can be finally obtained from (11).

It must be noted that, from the relation $a(\tau)$, it is also possible to obtain the speed-up factor, $\xi \equiv H/H^{FRW}$, where $H \equiv \dot{a}/a$ is the Hubble parameter, while $H^{FRW}$ is that predicted by GR at the same temperature. To that end, we use the relation $H^{FRW} \propto a^{-2}$ together with Eq. (11), and we find

$$\xi \propto a' \qquad (19)$$

where $\xi$ can be normalized so that $\xi \to 1$ as $\tau \to +\infty$.

For future purposes, it is convenient to write Eq. (17) in a more compact form. To that end, we define

$$p = \frac{1}{2} \ln |y| \qquad (20)$$

and

$$\varphi = -\frac{1}{2} \int |3 + 2\omega|^{1/2} \frac{d\phi}{\phi} \qquad (21)$$

By using these quantities, Eq. (17) can be then written as (with $K \equiv -A\Gamma^{1/2}$ and $C = \text{sign}(K)\pi/2$)

$$\varphi = \begin{cases} -\sqrt{3} \ln[Ke^{-p} + (1 + K^2 e^{-2p})^{1/2}] & (3 + 2\omega > 0) \\ -\sqrt{3} \left( \arctan \left[ \frac{(1 - K^2 e^{-2p})^{1/2}}{Ke^{-p}} \right] - C \right) & (3 + 2\omega < 0) \end{cases} \qquad (22)$$

and, hence

$$\frac{d\varphi}{dp} = -\frac{1}{2} |3 + 2\omega|^{1/2} \frac{1}{\phi} \frac{d\phi}{dp} \qquad (23a)$$

$$= \begin{cases} \sqrt{3} K (e^{2p} + K^2)^{-1/2} & (3 + 2\omega > 0) \\ -\sqrt{3} K (e^{2p} - K^2)^{-1/2} & (3 + 2\omega < 0) \end{cases} \qquad (23b)$$

It is interesting to note that the function $\varphi$ defined in Eq. (21) corresponds (except for a constant term) to the Einstein scalar field given by Eqs. (3) and (4). Similarly, Eq. (22) for $3 + 2\omega > 0$ coincides with the radiation-dominated solution obtained by Damour & Nordtvedt [23] by using the Einstein frame. However, it is important to remark that we are always using the physical Jordan frame. We just use the quantities $\varphi$ and $p$ as an abbreviated notation allowing an easy comparison of some of our results and those obtained by using the Einstein frame.

## III. ASYMPTOTIC BEHAVIORS OF SCALAR-TENSOR THEORIES: CONVERGENCE TO GENERAL RELATIVITY

We are only interested on viable scalar-tensor cosmological models, that is, those which are compatible with present solar-system observations but where the early evolution of the Universe is distinguishable from that obtained in GR.

In order to ensure compatibility with solar-system experiments, we will later consider that any viable scalar-tensor gravity theory must converge towards GR ($|\omega| \to \infty$ and $|\omega'/\omega^3| \to 0$) as $\tau \to +\infty$ (equivalent to $t \to \infty$). We show elsewhere [43] that primordial nucleosynthesis requires in fact that scalar-tensor theories which converge towards GR are "indistinguishable" from GR as soon as **the end** of the radiation-dominated era.

A large bound on the present value of $\omega$ (hereafter $\omega_0$) implies [3] that

$$\Phi_0 = \frac{4 + 2\omega_0}{3 + 2\omega_0} \simeq 1 \tag{24}$$

where $\Phi_0$ is the present value of the dimensionless scalar field

$$\Phi \equiv G_0 \phi. \tag{25}$$

Consequently, if we define $x = |\Phi - 1|$, the asymptotic behavior of $|3 + 2\omega|$ when $\tau \to +\infty$ ($x \to 0$) can be described as some power law $3\lambda^{-2} x^{-\epsilon}$ (with $\lambda > 0$ and $\epsilon > 0$). We have used $3\lambda^{-2}$ for the proportionality constant to facilitate the comparison with the Eq. (53) below. It is known [39] that convergence towards GR ($|\omega'/\omega^3| \to 0$) requires $\epsilon > 1/2$. We will show now that compatibility with solar-system experiments also requires an upper bound on $\epsilon$. To that end, we will distinguish the case in which $\Phi$ converges towards the GR value ($\Phi = 1$) as $\Phi \to 1^-$, from the case in which $\Phi \to 1^+$.

If $\Phi \to 1^-$ ($x = 1 - \Phi$), the integral $\varphi$ appearing in Eq. (17) and (21) can be then evaluated in a convenient closed form for a variety of values of $\epsilon$. When $\epsilon/2$ is a half-integer, we have

$$\varphi = -\frac{\sqrt{3}}{2\lambda}\left(\ln\left[\frac{1-\sqrt{x}}{1+\sqrt{x}}\right] + \sum_{k=1}^{(\epsilon-1)/2} \frac{2x^{-(2k-1)/2}}{2k-1}\right) \tag{26}$$

and when $\epsilon/2$ is an integer

$$\varphi = -\frac{\sqrt{3}}{2\lambda}\left(\ln\left[\frac{1-x}{x}\right] + \sum_{k=1}^{(\epsilon-2)/2} \frac{x^{-k}}{k}\right) \tag{27}$$

In the same way, if $\Phi \to 1^+$ ($x = \Phi - 1$), the integration of Eq. (21) for half-integral values of $\epsilon/2$ gives

$$\varphi = -\frac{\sqrt{3}}{2\lambda}(-1)^{\frac{\epsilon+3}{2}}$$
$$\times \left\{2\arctan(\sqrt{x}) + \sum_{k=1}^{(\epsilon-1)/2} (-1)^{k+1}\frac{2x^{-(2k-1)/2}}{2k-1}\right\} \tag{28}$$

while, for integral values of $\epsilon/2$

$$\varphi = -\frac{\sqrt{3}}{2\lambda}(-1)^{\frac{\epsilon+2}{2}}$$
$$\times \left\{\ln\left(\frac{x}{x+1}\right) + \sum_{k=1}^{(\epsilon-2)/2} (-1)^{k+1}\frac{x^{-k}}{k}\right\} \tag{29}$$

From Eq. (22), we see that $\varphi \to 0$ as $p \to +\infty$ whatever the sign of $(3 + 2\omega)$ is. This condition can only be reconciled with $x \to 0$, as required by solar-system experiments, if $\epsilon$ (integer or half-integer) is strictly smaller than 2 in the expressions (26)–(29). As a matter of fact, if $\epsilon \geq 2$, the solutions (26)–(29) are dominated by the terms with the form $x^{-\alpha}$, which diverge to infinity as $x \to 0$. It is important to notice that in the case analyzed by Barrow [39], ($\epsilon = 2$, $\Phi < 1$ and $3 + 2\omega > 0$), the limit $\varphi \to 0$ implies $x \to 1/2$. Barrow identified this solution with that obtained in GR because $\Phi \to$const ensures $\omega(\Phi) \to$const. However, we note that the constant value to which $\Phi$ approaches is $1/2$ and Eq. (24) then implies that the value to which $\omega$ approaches is not compatible with solar-system experiments.

By generalizing the previous analytical results to non-integer and non-half-integer $\epsilon$ values (we have tested the validity of such generalization by means of numerical computations), we then conclude that convergence towards GR as $\tau \to +\infty$, requires

$$1/2 < \epsilon < 2 \qquad (30)$$

## IV. QUALITATIVE BEHAVIORS OF SCALAR-TENSOR THEORIES AT EARLY TIMES

If $\Phi$ becomes equal to unity ($\varphi = 0$) at a finite time, the theory is GR at all the radiation-dominated times, because in this case Eq. (22) implies $K = 0$ and then $\varphi$ is always vanishing. Moreover, in that case, Eq. (23a) implies that $\omega$ is infinite. However, an infinite $\omega$ is not equivalent to have GR, excepted if $\omega(\Phi)$ is monotonic. Since we consider only monotonic $\omega(\Phi)$ and we are only interested in models where the early evolution of the universe is distinguishable from that obtained in GR, we will consider that $3 + 2\omega$ must be finite at $\tau = 0$. Its value can be however positive, vanishing or negative. We will now analyze separately these three possible cases by using the method described in Sect. II B.

The criteria we have used to set the origin of $\tau$-times are the following: when models are singular, that is, when $a(\tau)$ vanishes at some $\tau$, we take the constant $\tau_0$ appearing in Eq. (16) so that we have the more recent singularity at $\tau = 0$. On the contrary, if the scale factor has some non-vanishing minimum, the origin of times will be taken so that there is only one minimum $a$-value for positive $\tau$ values.

### A. Possible values of $\Phi$ at $\tau = 0$.

In order to represent adequately the early behavior of $3 + 2\omega$ we need to know which are the possible values of $\Phi$ at $\tau = 0$. We will now show that simple arguments considerably reduce the range of possible values of $\Phi(0)(= \Phi(\tau = 0))$.

#### 1. Case $3 + 2\omega$ positive at $\tau = 0$

If $3 + 2\omega > 0$ at $\tau = 0$, $\Phi(0)$ must be $+\infty$ or $0$. The other possibilities for $\Phi(0)$ are excluded.

Since $3+2\omega(\tau = 0)$ is finite, if $\Phi(0)$ diverges to $-\infty$, then $3+2\omega(\Phi)$ (as a function of $\Phi$) has a nonvanishing horizontal asymptote. The early behavior of the coupling function could be then described by $(3 + 2\omega) \simeq |k|$ at $\tau \to 0$, and Eqs. (17) would imply $\Phi = -f\sqrt{3/k}$. Consequently, $a^2 \equiv y\Phi^{-1} < 0$ and the dimensionless scale factor would become imaginary.

On the other hand, if $\Phi(0)$ has some nonvanishing finite value, we have a contradiction with the limit $d\varphi/dp \to \sqrt{3}$ as $p \to -\infty$ implied by Eq.(23b). As a matter of fact, in such a case we can choose $\tau_0 = A$ in Eq. (16) so that, from Eq.(12), the scale factor vanishes at $\tau = 0$. From Eq.(20), we then deduce $p \to -\infty$ as $\tau \to 0$. The resulting cosmological models are singular and $\Phi(p)$ has a nonvanishing horizontal asymptote at early times. Consequently, $d\Phi/dp \to 0$ and, from Eq.(23a), $d\varphi/dp \to 0 \neq \sqrt{3}$.

#### 2. Case $3 + 2\omega$ vanishing at $\tau = 0$

If $3+2\omega(\tau = 0) \to 0$, then $\Phi(0)$ must diverge to $\pm\infty$ (if $3+2\omega \to 0^-$ the case $\Phi(0) \to -\infty$ is excluded) or vanish.

From the same arguments as before we can exclude any other possibility for $\Phi(0)$. In the case where $\Phi(0)$ diverges to $-\infty$ and $3 + 2\omega(\tau = 0) \to 0^-$ (from negative values), Eq. (17) would lead to solutions with $a^2 < 0$. Nonvanishing finite values of $\Phi(0)$ would lead to values of $d\varphi/dp$ at early times which are not compatible with that obtained from Eq. (23b).

### 3. Case $3 + 2\omega$ negative at $\tau = 0$

If $3 + 2\omega(\tau = 0) < 0$, then $\Phi(0)$ must be positive and finite.

The non-negative value of $\Phi(0)$ comes from the condition $a^2 \equiv y\Phi^{-1} \geq 0$ where, from Eq. (16), $y$ is always positive when $3 + 2\omega < 0$. On the other hand, $\Phi(0)$ must be finite because, otherwise, $3 + 2\omega(\Phi)$ would have a nonvanishing horizontal asymptote as $\Phi \to \infty$. The early behavior of the coupling function could be then described by $-(3 + 2\omega) \simeq |k|$ and Eq. (21) would imply $\varphi \propto \ln(\Phi) \to \infty$ as $\tau \to 0$, in contradiction with the finite value of $\varphi$ implied by Eq. (22) when $3 + 2\omega < 0$.

## B. Solutions at early times

In order to analyze all possible early behaviors of scalar-tensor theories, the coupling function near $\tau = 0$ will be expressed as a power law of $x^{-1}$ ($x = |\Phi - 1|$). Sometimes it will be necessary to add a constant $k$ to get the assumed $\omega$ value at $\tau = 0$. This additive constant also allows an easier comparison with our analysis of section V. The negative power of $x$ allows for larger convergence radii of our expressions.

The analytical expressions of all the possible early behaviors are now listed.

### 1. Case $3 + 2\omega$ positive at $\tau = 0$

- If $3 + 2\omega > 0$ as $\Phi \to +\infty$, the asymptotic behavior of the coupling function at early times can be described by

$$3 + 2\omega \simeq (3/\lambda^2)(k + x^{-1}) \qquad (\lambda > 0, k > 0, x = \Phi - 1). \tag{31}$$

Integration of Eq. (17) then leads to

$$2\sqrt{1-k}\arctan\left(\frac{\sqrt{1-k}}{\sqrt{x^{-1}+k}}\right) - \sqrt{k}\ln\left(\frac{\sqrt{x^{-1}+k}-\sqrt{k}}{\sqrt{x^{-1}+k}+\sqrt{k}}\right) = \ln(f^\lambda)(0 \leq k \leq 1) \tag{32}$$

$$\sqrt{k-1}\ln\left(\frac{\sqrt{x^{-1}+k}-\sqrt{k-1}}{\sqrt{x^{-1}+k}+\sqrt{k-1}}\right) - \sqrt{k}\ln\left(\frac{\sqrt{x^{-1}+k}-\sqrt{k}}{\sqrt{x^{-1}+k}+\sqrt{k}}\right) = \ln(f^\lambda) \quad (k \geq 1)$$

$$\tag{33}$$

where $f$ is given by (18). At early times ($x \to +\infty$), the dominant term in the left-hand side of Eqs. (32) is the second one, whatever the value of $k$ is, and therefore

$$a \simeq 2\sqrt{\Gamma G_0 k}\tau \left(\frac{f^{\lambda/\sqrt{k}-1}}{1+2(2k-1)f^{\lambda/\sqrt{k}}+f^{2\lambda/\sqrt{k}}}\right)^{1/2} \to \propto \tau^{(1+\lambda/\sqrt{k})/2} \propto t^{\frac{1+\lambda/\sqrt{k}}{3+\lambda/\sqrt{k}}}$$

$$\Phi \simeq 1 + \frac{1}{4k}(f^{\lambda/\sqrt{k}} + f^{-\lambda/\sqrt{k}} - 2) \to \propto \tau^{-\lambda/\sqrt{k}} \propto a^{\frac{-2\lambda/\sqrt{k}}{1+\lambda/\sqrt{k}}} \tag{34}$$

$$\xi \to \propto \tau^{(\lambda/\sqrt{k}-1)/2} \propto a^{\frac{\lambda/\sqrt{k}-1}{\lambda/\sqrt{k}+1}}$$

where we have again taken $\tau_0 = A$.

We see that all models described by Eq. (34) are singular ($a(0) = 0$), but the $\xi$ value at $\tau = 0$ is: *i)* $\xi(0) \to +\infty$, if $\lambda/\sqrt{k} < 1$; *ii)* $\xi(0) \to$ const., if $\lambda/\sqrt{k} = 1$; or *iii)* $\xi(0) \to 0$, if $\lambda/\sqrt{k} > 1$.

This class of solutions is the most often studied in the literature [37,40]. It can be considered as a perturbed constant-$\omega$ Brans-Dicke theory. As in that particular theory, models have a power-law behavior at early times. However, since $\omega(\phi)$ has not been taken as a constant, the solar-system constraints on $\omega_0$ do not hold at early times. Consequently, this class of theories is specially well suited to be used in extended inflation models. Our analysis in terms of $a$, $\Phi$ and $\xi$ shows that, although all models have power law solutions at $\tau \to 0$, their speed-up factors can instead exhibit very different qualitative behaviors (see, e.g., the non-monotonic behavior of $\xi$ in theories defined by Eq. (31) with $\lambda/\sqrt{k} > 1$).

It is also important to note that the asymptotic solutions given by Eq. (34) represent a much wider family of theories than that directly defined by Eq. (31). For example, the case $3 + 2\omega \propto \Phi^2/(\Phi - 1)^2$ studied by Mimoso and Wands [40] has a Taylor approximation at early times ($\Phi \to +\infty$) given by $3 + 2\omega \propto 1/2 + x^{-1}$, which is a particular case of Eq. (31). The solutions found by those authors are in fact contained in Eq. (34) in the limit of small $\tau$ values.

- If $3 + 2\omega > 0$ as $\Phi(0) \to 0$. We can describe this case by taking

$$3 + 2\omega \simeq (3/\lambda^2)(k + x^{-1}) \qquad (\lambda > 0, k > -1, x = 1 - \Phi). \qquad (35)$$

Equation (17) then leads to

$$\sqrt{k+1} \ln\left(\frac{\sqrt{x^{-1}+k} - \sqrt{k+1}}{\sqrt{x^{-1}+k} + \sqrt{k+1}}\right) - 2\sqrt{-k} \arcsin(\sqrt{-kx}) = \ln(f^\lambda) (-1 \leq k \leq 0) \quad (36)$$

$$\sqrt{k+1} \ln\left(\frac{\sqrt{x^{-1}+k} - \sqrt{k+1}}{\sqrt{x^{-1}+k} + \sqrt{k+1}}\right) - \sqrt{k} \ln\left(\frac{\sqrt{x^{-1}+k} - \sqrt{k}}{\sqrt{x^{-1}+k} + \sqrt{k}}\right) = \ln(f^\lambda) \quad (k \geq 0)$$

where, in the limit $\Phi \to 0$ ($x \to 1$), the dominant term in the left-hand side is the first one, whatever the value of $k$ is. Then

$$a \simeq \sqrt{\Gamma G_0/(4(k+1))}\tau \left(f^{-(1+\lambda/\sqrt{k+1})}(1+2(2k-1)f^{\lambda/\sqrt{k+1}}+f^{2\lambda/\sqrt{k+1}})\right)^{1/2}$$
$$\to \propto \tau^{(1-\lambda/\sqrt{k+1})/2} \propto t^{\frac{1-\lambda/\sqrt{k+1}}{3-\lambda/\sqrt{k+1}}}$$
$$\Phi \simeq \frac{4(k+1)f^{\lambda/\sqrt{k+1}}}{1+2(2k+1)f^{\lambda/\sqrt{k+1}}+f^{2\lambda/\sqrt{k+1}}} \to \propto \tau^{\lambda/\sqrt{k+1}} \propto a^{\frac{2\lambda/\sqrt{k+1}}{1-\lambda/\sqrt{k+1}}} \quad (37)$$
$$\xi \to \propto \tau^{-(1+\lambda/\sqrt{k+1})/2} \propto a^{\frac{\lambda/\sqrt{k+1}+1}{\lambda/\sqrt{k+1}-1}}$$

where we have taken $\tau_0 = A$ to get $\tau = 0$ at the singularity. Arrows in the previous expressions denote the limit $\tau \to 0$.

We see from these equations that the condition $\Phi \to 0$ as $\tau \to 0$ requires: *i)* $0 < \lambda/\sqrt{k+1} < 1$, or *ii)* $1 < \lambda/\sqrt{k+1} < 3$. In the first case, models are singular and $\xi(0) \to +\infty$ while, in the second case, models are nonsingular and $\xi(0) \to -\infty$.

The particular case $k = 0$ of Eq. (35) has been previously considered by Barrow [39] and Mimoso and Wands [40], who found analytical solutions which are consistent with those given by Eq. (37).

We have also performed some similar computations using other possible representations for $\omega(\Phi)$ as, for example, $3 + 2\omega = (3/\lambda^2)x^{\pm\epsilon}$. The general solutions of Eq. (17) can then have a form which is different from that given by Eq. (37) but, in the limit $\tau \to 0$, all expressions become identical ($a \propto \tau^{(1-\lambda)/2}$, $\Phi \propto \tau^\lambda$, $\xi \propto \tau^{-(1+\lambda)/2}$).

### 2. Case $3 + 2\omega$ vanishing at $\tau = 0$

- If $3 + 2\omega(0) \to 0$ and $\Phi(0) \to +\infty$, we can analyze this case by considering

$$|3+2\omega| \simeq (3/\lambda^2)x^{-\epsilon} \qquad (\epsilon > 0, x = \Phi - 1). \quad (38)$$

The integral appearing in Eq. (17) has then the solutions given in Eqs. (28)-(29). However, since now $x \to +\infty$, the dominant terms in these solutions are $(-1)^{(\epsilon+3)/2}2\arctan(\sqrt{x})$ when $\epsilon/2$ is a half-integer, and $(-1)^{(\epsilon+2)/2}\ln[x/(x+1)]$ when $\epsilon/2$ is an integer in Eq. (28) and Eq. (29) respectively. It must be noted that only the solutions of Eq. (17) in the case where $\epsilon/2$ is an half-integer satisfy the condition $\Phi \to +\infty$ defining this class of asymptotic models. Consequently, the case where $\epsilon/2$ is integer will be not included here.

If $3 + 2\omega(0) \to 0^+$, Eqs. (17), (28) and (29) lead to

$$a \simeq \sqrt{\Gamma G_0/2}[(\tau + \tau_0)^2 - A^2]^{1/2} \, |\cos(\ln(f^{\lambda/2}))| \to 0$$
$$\Phi \simeq \cos^{-2}[\ln(f^{\lambda/2})] \to +\infty \tag{39}$$
$$\xi \to A\frac{\lambda}{2}(e^{\pi/\lambda} - 1)/e^{\pi/2\lambda} = const.$$

where $f$ is given by Eq. (18), with $\tau_0$ taken as $\tau_0 = A(e^{\pi/\lambda} + 1)/(1 - e^{\pi/\lambda})$ in order to have the more recent singular point at $\tau = 0$.

From Eq. (39), we see that all these models are singular, with $\xi(0) \to const$. A particular case of this class of behaviors is Barker's theory, defined by $3 + 2\omega = x^{-1}$ ($\lambda^2 = 3$ and $\epsilon = 1$).

On the other hand, if $3 + 2\omega(0) \to 0^-$, we obtain

$$a = \sqrt{2\Gamma G_0} A(1 + [(\tau + \tau_0)/A]^2)^{1/2} \cos(g) \to \propto \tau \propto t^{1/2}$$
$$\Phi = \cos^{-2}(g) \to \propto \tau^{-2} \propto a^{-2} \tag{40}$$
$$\xi \to (\lambda/\sqrt{2})[1 - \cos(\pi/\lambda)]^{1/2}$$

where $g = \lambda[\arctan[(\tau + \tau_0)/A] - \pi/2]$ and we have taken $\tau_0 = A\cot(\pi/2\lambda)$ (with $\lambda \geq 1/2$) to satisfy the condition $\Phi(0) \to +\infty$ (or $\cos(g) \to 0$). This condition cannot be obtained for $\lambda < 1/2$ and, hence, such a range of $\lambda$ values is excluded here.

Models are again singular, with $\xi \to const.$, where the constant is greater than unity for $\lambda > 1$ and smaller than unity for $\lambda < 1$.

- If $3 + 2\omega(0) \to 0^+$ and $\Phi(0) \to -\infty$, we can describe the asymptotic form of the coupling function by taking

$$(3 + 2\omega) \simeq (3/\lambda^2)x^{-\epsilon} \qquad (\epsilon > 0, x = 1 - \Phi). \tag{41}$$

The integral appearing in Eq. (17) has then the solutions given in Eqs. (26) and (27) but, since $x \to \infty$, the dominant terms are now the logarithmic ones. For half-integral values of $\epsilon/2$, we obtain

$$a = \sqrt{\Gamma G_0 \tau} \to \propto \tau \propto t^{1/2}$$
$$\Phi = \frac{4f}{(1+f)^2} = \frac{\tau^2 - A^2}{\tau^2} \to \propto -\tau^{-2} \propto -a^{-2} \qquad (42)$$
$$\xi = 1$$

where we have taken $\tau_0 = 0$ in order to obtain $\tau = 0$ at the singularity, and $\lambda = 1$ in order to satisfy the condition $\Phi(0) \to -\infty$.

We see that the expansion of the universe is equal (at any time) to that obtained in GR. This result can be understood by using Eqs. (42) and the field equations (9) to verify that the scale factor, $a$, then satisfies the standard Friedman-Robertson-Walker equation, while the scalar-field satisfies an uncoupled differential equation.

A similar behavior is found for integral values of $\epsilon/2$.

- If $3 + 2\omega(0) \to 0$ and $\Phi(0) \to 0$, we can represent this case by taking

$$|3 + 2\omega| \simeq (3/\lambda^2)(x^{-1} - 1) \qquad (x = 1 - \Phi). \qquad (43)$$

Then, if $3 + 2\omega \to 0^+$, Eq. (17) implies

$$-2 \arcsin \sqrt{x} = \ln(f^\lambda) \qquad (44)$$

and, then

$$a = \sqrt{\Gamma G_0} \frac{[(\tau + \tau_0)^2 - A^2]^{1/2}}{\cos(\ln f^{-\lambda/2})}$$
$$\to \propto \cos^{-1}(\ln f^{-\lambda/2}) \to \infty$$
$$\Phi = \cos^2(\ln f^{-\lambda/2}) \to 0 \qquad (45)$$
$$\xi \to \propto -\cos^{-2}(\ln f^{-\lambda/2}) \to -\infty$$

where $\tau_0 = A(e^{\pi/\lambda} + 1)/(e^{\pi/\lambda} - 1)$ in order to have $\Phi(0) \to 0$ which defines this case. This choice of $\tau_0$ implies that $a(0) \to +\infty$ and models are then nonsingular, with a nonvanishing minimum value of $a$ at $\tau_{min} \simeq 2A/(e^{\pi/\lambda} - 1)$.

If $3 + 2\omega \to 0^-$, Eq. (17) gives

$$-\arcsin\sqrt{x} = \lambda\{\arctan[(\tau + \tau_0)/A] - \pi/2\} \qquad (46)$$

and we get

$$\begin{aligned} a &= \sqrt{\Gamma G_0}[(\tau + \tau_0)^2 + A^2]^{1/2}\,|\cos(g)|^{-1} \to \propto |\cos(g)|^{-1} \to +\infty \\ \Phi &= \cos^2(g) \to 0 \\ \xi &\to \propto -\cos^{-2}(g) \to -\infty \end{aligned} \qquad (47)$$

where $g = \lambda[\arctan(f) - \pi/2]$ and we have taken (for $\lambda \geq 1/2$) $\tau_0 = A\cot(\pi/2\lambda)$ in order to satisfy the condition $\Phi(0) \to 0$. We can see from Eq. (47) that these models are again nonsingular.

### 3. Case $3 + 2\omega$ negative at $\tau = 0$

- If $3 + 2\omega(0) < 0$ and $\Phi(0) < 1$, by taking

$$-(3 + 2\omega) \simeq (3/\lambda^2)x^{-1} \qquad (x = 1 - \Phi), \qquad (48)$$

Eq. (17) leads to

$$\ln\left[\frac{1 - \sqrt{x}}{1 + \sqrt{x}}\right] = 2\lambda\arctan(\tau + \tau_0)/A - \pi \qquad (49)$$

$$\begin{aligned} a &= \sqrt{\Gamma G_0/4}(\tau^2 + A^2)^{1/2}\frac{1 + e^g}{e^{g/2}} \to const \\ \Phi &= \frac{4e^g}{(1 + e^g)^2} \to const < 1 \\ \xi &= \frac{\tau}{2(\tau^2 + A^2)^{1/2}}\cdot\frac{(1 + e^g)}{e^{g/2}}\{1 + \frac{\lambda A}{\tau}[1 - \frac{2}{(1 + e^g)e^g}]\} \to -\infty \end{aligned} \qquad (50)$$

where $g = \lambda[\arctan[(\tau + \tau_0)/A] - \pi/2]$ with $\tau_0 = 0$.

We see that $a$ and $\Phi$ never vanish and models are then nonsingular.

- If $3 + 2\omega(0) < 0$ and $\Phi(0) > 1$, we can take $-(3 + 2\omega) \simeq (3/\lambda^2)x^{-1}$, with $x = |\Phi - 1| = \Phi - 1$. Solutions are then those given in Eqs. (40) but, now, with $\lambda \leq 1/2$ in order to have a finite $\Phi(0)$. In this case, $\cos(g)$ never vanishes and models are then nonsingular.

### C. Existence of an initial singularity

The existence of an initial singularity can be characterized in terms of the sign of $3+2\omega(0)$.

When $3 + 2\omega(0)$ is negative or vanishing ($\to 0^-$), we know that $\Phi(0)$ cannot diverge to infinity. Since the variable $y$ never vanishes when $3 + 2\omega(0) < 0$, Eq. (12) implies that all models are then nonsingular. The analysis of section IV B 3 shows that, in this case, the speed-up factor $\xi$ is negative at $\tau = 0$. The negative value of $\xi$ means a contraction phase of the Universe.

On the contrary, if $3+2\omega(0)$ is positive or vanishing ($\to 0^+$), we can choose $\tau_0$ in Eq. (16) so that $y \propto \tau \to 0$ at early times. From Eq. (12), we then find that only those models where $\Phi(0)\tau^\alpha \to 0$ ($\alpha > 1$) can be nonsingular. From Eqs. (45) and (37) we see that the condition $\alpha > 1$ is always satisfied when $3 + 2\omega(0) \to 0^+$ and $\Phi(0) \to 0$. However, if $3 + 2\omega(0) \to$ const, that condition requires $1 < \lambda/\sqrt{k+1} < 3$. As before, sections IV B 1 and IV B 2 show that the speed-up factor $\xi$ initially diverges to $-\infty$.

In all the other cases, models are singular and sections IV B 1 and IV B 2 show that the initial value of the speed-up factor $\xi$ can be 0, a positive constant, or $+\infty$.

Table 1 summarizes all the asymptotic solutions obtained in this Section as well as some of their main properties. Singular and non-singular models are denoted in the last column of this table by S and NS, respectively.

# V. RADIATION-DOMINATED EVOLUTION IN SCALAR-TENSOR COSMOLOGICAL MODELS

The evolution from the above asymptotic behaviors until their later convergence towards GR, can only be known by specifying a particular $\omega(\Phi)$, assumed to be valid during all that time interval. In order to get some insight into the possible evolution of scalar-tensor cosmological models, we will write $\omega(\Phi)$ as the following Taylor series.

$$\frac{1}{|\omega - b|^{1/\epsilon}} = \sum_{i=0}^{\infty} a_i \, |\Phi - 1|^i \qquad (51)$$

the first order of which is

$$\frac{1}{|\omega - b|^{1/\epsilon}} = a_1 \, |\Phi - 1| \qquad (52)$$

where, in order to allow for convergence towards GR, we have taken $a_0 = 0$ and $1/2 < \epsilon < 2$ (see Sect. III).

Eq. (52) is not just a generalization of that proposed by Garcia-Bellido and Quiros [22]. It gives an exact representation for most of the particular scalar-tensor theories proposed in the literature [44] and, what is more, it is also a first-order approximation to any other theory provided that $\Phi$ is not very different from unity. Solar-system experiments imply that this last condition is in fact satisfied from some time $t_*$ up to the present. For $t < t_*$, Eq. (52) then represents a particular class of theories which allows nevertheless to reproduce all the asymptotic behaviors described in the previous section.

A monotonic $\omega(\Phi)$ function, as that given by Eq. (52), implies that: *i)* $\Phi$ is never equal to unity except for $p \to +\infty$, and *ii)* $\Phi$ must have a monotonic time evolution. As a matter of fact, if $\Phi$ becomes equal to unity at a finite time, the theory is GR (i.e. $\varphi = 0$) at that time. Since, in that case, Eq. (22) implies $K = 0$, such a theory then reduces to GR at all the radiation-dominated times. On the other hand, if $\Phi$ is not a monotonic function of time, Eq. (23) would imply an infinite value of $\omega$ when $d\Phi/dp$ vanishes and, in turn, this would imply $\Phi = 1$. The theory then becomes identical to GR. Finally, since $\Phi$ cannot diverge to infinity except at $\tau = 0$ (because we then have $a^2 = y\Phi^{-1} \to 0$ and this defines the time $\tau = 0$), $\Phi$ is confined to values $\Phi > 1$, if the present value of $\Phi$ (hereafter $\Phi_0$) $\to 1^+$ or to $\Phi < 1$, if $\Phi_0 \to 1^-$.

In order to depict all possible cosmological models, we have then performed a numerical integration of the field equations (9) by using Eq. (52) and the temperature $T$ as variable. These numerical solutions can be easily interpreted by considering the asymptotic behaviors found in previous sections. In order to facilitate the discussion, it is convenient to rewrite the function $3 + 2\omega$ appearing in Eq. (17) as

$$|3 + 2\omega| = (3/\lambda^2)(x^{-\epsilon} + k) \qquad (53)$$

where $\lambda \equiv \sqrt{3 |a_1|^\epsilon / 2}$, $k \equiv (\lambda^2/3)(3 + 2b)$, and

$$x = \begin{cases} 1 - \Phi & (\text{if } \Phi < 1) \\ \Phi - 1 & (\text{if } \Phi > 1) \end{cases} \qquad (54)$$

where the condition $\Phi_0 < 1$ is obtained for $a_1 < 0$ and $3+2\omega_0 > 0$, or $a_1 > 0$ and $3+2\omega_0 < 0$, while $\Phi_0 > 1$ is obtained for $a_1 > 0$ and $3 + 2\omega_0 > 0$, or $a_1 < 0$ and $3 + 2\omega_0 < 0$.

We will now discuss the different cosmological models in terms of the $k$ and $\lambda$ parameters. The qualitative behaviors of these models does not depend on $\epsilon$ provided that $1/2 < \epsilon < 2$.

### A. Models with $\omega_0 \gg 0$

#### 1. Case $\Phi_0 \to 1^-$ and $k > -1$

In this case, Eq. (53) implies that $3 + 2\omega$ is strictly positive at any time and that $\Phi$ remains always smaller than unity. According to that quoted in section IV A, $\Phi$ vanishes at $\tau = 0$. The early evolution of the universe then corresponds to the asymptotic solutions of section IV B 1 (Eq. (37)) and depends on the $\lambda/\sqrt{k+1}$ value.

- If $0 < \lambda/\sqrt{k+1} < 1$, cosmological models are singular. $\xi$ decreases monotonically from $+\infty$ at $\tau = 0$ to its present value $\to 1^+$ (Figure 1).

- If $1 < \lambda/\sqrt{k+1} < 3$, cosmological models are nonsingular. The scale factor has a minimum value at some time $\tau_{min}$, where $\xi = 0$, and the temperature of the universe has then a maximum value ($T(\tau_{min})$ (Figure 2).

  If $k = 0$ and $\epsilon = 1$, the neglected terms in Eq. (36) are exactly vanishing. Hence, the analytical solutions given by Eqs. (37) are exact at any time. We have therefore found nonsingular models which, unlike those obtained by Barrow [39], converge towards GR as $\tau \to +\infty$ and, hence, they are compatible with all solar-system experiments.

- If $k = 0$ and $\lambda = 1$, solutions are those given by Eq. (42) and the scalar field is then uncoupled from the expansion of the Universe ($\xi$ is then equal to unity at any time).

*2. Case $\Phi_0 \to 1^+$ and $k > 0$*

In this case, Eq. (53) again implies that $3 + 2\omega$ is strictly positive at any time and $\Phi(0) \to +\infty$. The early evolution of these models then corresponds to the asymptotic solutions given by Eq. (34).

Models are singular, with $\Phi$ monotonically decreasing from $\Phi(0) \to +\infty$. The behavior of $\xi$ depends instead on the value of $\lambda/\sqrt{k}$:

- If $\lambda/\sqrt{k} < 1$, $\xi$ decreases monotonically at any time until reaching unity (Figure 3).

- If $\lambda/\sqrt{k} = 1$, $\xi(0) \to const$. Numerical integration shows in this case that the finite value of $\xi(0)$ is always greater than 1 (Figure 4).

- If $\lambda/\sqrt{k} > 1$, then $\xi$ initially increases from $\xi(0) \to 0$ and later reaches a maximum value $\xi_{max}$ which can be arbitrarily high. The $\xi_{max}$ value increases with $k$, while it slightly decreases with $\lambda$. The radiation-dominated behavior of $\xi(\tau)$ is then highly non-monotonic, with an initial phase where $\xi$ is slower than in GR and a later phase with a faster expansion of the universe which converges towards the GR value as $\tau \to +\infty$ (Figure 5). No similar behaviors have been found for any other choice of parameters, even for negative $\omega_0$ values. We have not either found models where $\xi$ has some minimum value at $\tau > 0$.

### 3. Case $\Phi_0 \to 1^-$ and $k = -1$

$3+2\omega$ is always positive and vanishes at $\tau \to 0$, with $\Phi(0) \to 0$. The radiation-dominated evolution is then described by Eq. (45), which are exact solutions for $\epsilon = 1$, and models are therefore nonsingular. The qualitative behavior of these models is then similar to that found in section V A 1 when $1 < \lambda/\sqrt{k+1} < 3$. However, in this case, the speed-up has a smooth maximum value greater than unity after of which it falls quickly to zero (Figure 6). Note that solutions given by Eq. (45) provide another class of nonsingular models compatible with solar-system experiments.

### 4. Case $\Phi_0 \to 1^+$ and $k = 0$

$3 + 2\omega$ is always positive and vanishes at $\tau \to 0$, with $\Phi \to +\infty$. Solutions at any time are those given by Eqs. (28) and (29) while, at early times, they are approximated by Eqs. (39). Their early behavior is then similar to that discussed in section V A 2 ($\lambda/\sqrt{k} = 1$) (Figure 4). These models are then singular, with $\Phi(0) \to +\infty$ and $\xi \to const > 1$.

### 5. Cases $\Phi_0 \to 1^-$ and $k < -1$ or $\Phi_0 \to 1^+$ and $k < 0$

These values of the parameters in Eq. (53) imply that $3 + 2\omega$ vanishes at some finite temperature, where $\Phi = \Phi_*$. Eq. (23) then implies that $d\Phi/dp \to -\infty$ as $\Phi \to \Phi_*$ and, from Eqs. (12) and (19), we have also that $\xi \to +\infty$ in this limit. This class of theories is very unsatisfactory from a theoretical point of view. As a matter of fact, if the sign of $3 + 2\omega$ changes at a finite temperature, it is not possible to set the integration constants in Eqs. (22) and (23) to avoid a discontinuity in $\varphi$ or $d\varphi/dp$. Numerical integration produces in fact overflows at this point (because of the infinite values of $a'$, $\Phi'$ and $(3+2\omega)^{-1}$). We have not tried to handle this limit because we consider that such a change in the sign of $3 + 2\omega$ is not physical. Numerical integration stops then at such a temperature (see Figure 7).

### B. Models with $\omega_0 \ll 0$

#### 1. Case $\Phi_0 \to 1^-$ and $k < 1$ or $\Phi_0 \to 1^+$ and $k < 0$

In the case $\Phi_0 \to 1^-$ and $k < 1$ ($\Phi_0 \to 1^+$ and $k < 0$), $\Phi(0)$ has some nonvanishing value smaller (greater) than unity and $3 + 2\omega$ grows (decreases) converging towards $(3/\lambda^2)(k - |1 - \Phi(0)|^{-1})$ at $\tau \to 0$. Numerical integrations show however that $\Phi(0)$ is very small (large) and, therefore, $3 + 2\omega(0)$ has some nonvanishing negative value. The early behavior is then similar to that discussed in section IV B 3 (Eq. [50]) for $\Phi < 1$ ($\Phi > 1$). Consequently, models are nonsingular and their qualitative behavior is similar to that shown in Figure 2.

#### 2. Case $\Phi_0 \to 1^-$ and $k = 1$

In this case, $\Phi(0) \to 0$, $3 + 2\omega$ is negative at any time, and it approaches $0^-$ at $\tau = 0$. The asymptotic behavior of this model is given by Eq. (47). As before, it is nonsingular and the evolution of $\xi$ is that of Figure 2.

#### 3. Case $\Phi_0 \to 1^+$ and $k = 0$

In this case, $\Phi(0) \to +\infty$, $3 + 2\omega$ is negative at any time, and it approaches zero as $\tau \to 0$. The early evolution of $\xi$ is given by Eq. (40) and depends on the $\lambda$ parameter:

- If $\lambda > 1$, then $\xi(0) > 1$, and the qualitative behavior is similar to that of section V A 2 for $\lambda/\sqrt{k} = 1$ (see Figure 4).

- If $\lambda = 1$, then $\xi(0) = 1$. Numerical integration shows that $\xi = 1$ at any time and, therefore, the scalar-field is uncoupled from the expansion of the universe. The behavior of these models is then equal to that of GR (see section V A 1).

- If $\lambda < 1$, then $\xi(0) < 1$, and the qualitative behavior is that shown in Figure 8. These are the only models we have found with a non-vanishing value of $\xi(0)$ smaller than 1.

*4. Cases $\Phi_0 \to 1^-$ and $k > 1$ or $\Phi_0 \to 1^+$ and $k > 0$*

These models imply that $3 + 2\omega$ vanishes at some finite temperature. Their behavior has been described in section V A 5 (Figure 7).

## VI. CONCLUSIONS

We have analytically and numerically studied all the possible behaviors of scalar-tensor cosmologies with an arbitrary monotonic $\omega(\Phi)$ function. Our main results are:

1. The convergence towards GR during radiation-dominated era, requires an asymptotic behavior of the coupling function as $\mid 3 + 2\omega \mid \propto \mid 1 - \Phi \mid^{-\epsilon}$, with $1/2 < \epsilon < 2$. On the other hand, Damour & Nordtvedt [23] have found, by using the Einstein frame, a natural attractor mechanism toward GR during the matter-dominated era. Although the nature of these two conditions are totally different, it is straightforward to show that our condition is mathematically equivalent to that of Damour & Nordtvedt. Consequently, any model which does not satisfy our condition on the $\epsilon$ parameter, will not either converge toward GR during the matter-radiation era.

2. We have shown that the value of the $\Phi$ field at $\tau = 0$ and the existence of an initial singularity are constrained by the sign of $3 + 2\omega$ at that time. In particular, if $3 + 2\omega \geq 0$, $\Phi(0)$ cannot have a nonvanishing finite value. On the contrary, if $3 + 2\omega < 0$, $\Phi$ must be finite and all the resulting models are non-singular.

3. The analytical asymptotic solutions of all the possible scalar-tensor cosmological models converging towards GR have been discussed in terms of the sign of $3+2\omega$ at $\tau = 0$. These solutions describe all the early behaviors for any monotonic form of the coupling function $\omega(\Phi)$. Although some of these solutions are known in the literature[1] (e.g., [28,39,40]), we have been able to find several other behaviors which are qualitatively very different from the previous ones. We remark in particular that several of these solutions imply non-singular models which, unlike that found by Barrow [39], are compatible with solar-system experiments.

4. From a specified form of the coupling function $\omega(\Phi)$, which allows nevertheless to reproduce all the above asymptotic behaviors, we have numerically determined the evolution of the universe at any time. Our analysis reveals the existence of scalar-tensor cosmological models with qualitative behaviors very different from those previously found in the particular cases studied in the literature. We summarize in Figures 9 all these behaviors in terms of the $\lambda$ and $k$ parameters defined in Eq. (53). It is important to note that, for the first time, we have found a class of models with an explicit non-monotonic time evolution of the speed-up factor.

Finally, additional constraints on all these behaviors can be obtained from the light elements primordial production. This question is examined in a forthcoming paper.

### ACKNOWLEDGMENTS

We thank T. Damour and G. Esposito-Farese for useful discussions about the formulation of Scalar-Tensor theories by using the Einstein Frame. We are also grateful to L. Blanchet and D. Langlois for very valuable comments about this work.---

[1] Except for that concerning the $\xi(\tau)$ solutions

FIGURES

FIG. 1. Singular models with $\Phi(0) = 0$ and $\xi$ decreasing monotonically from infinity ($\lambda^2 = 3/4$, $k = 0$, solid line corresponds to $\omega_0 = 10^{22}$ and dashed line to $\omega_0 = 10^{24}$). Note that the largest $\omega_0$ values the most shifted to the left are the curves. However, the form of curves does not depend on $\omega_0$

FIG. 2. Non-singular models with $\xi$ increasing from 0 ($\lambda^2 = 3/2$, $k = 0$, solid line corresponds to $\omega_0 = 10^{23}$ and dashed line to $\omega_0 = 10^{25}$)

FIG. 3. Singular models with $\Phi(0) \to +\infty$ and $\xi$ decreasing monotonically $\lambda^2 = 3/2$, $k = 26.5$, solid line corresponds to $\omega_0 = 10^{20}$ and dashed line to $\omega_0 = 10^{22}$)

FIG. 4. Singular models with $\Phi(0) \to +\infty$ and $\xi$ decreasing from a finite constant value, $\xi(0) > 1$ ($\lambda^2 = 3/2$, $k = 3/2$, solid line corresponds to $\omega_0 = 10^{18}$ and dashed line to $\omega_0 = 10^{20}$).

FIG. 5. Singular models with $\Phi(0) \to +\infty$ and a highly non-monotonic $\xi$ ($\lambda^2 = 3/2$, $k = 1/2$, solid line corresponds to $\omega_0 = 10^{18}$ and dashed line to $\omega_0 = 10^{20}$)

FIG. 6. Nonsingular models with $\Phi(0) = 0$ and $\xi$ has a smooth maximum value greater than unity ($\lambda^2 = 3/4$, $k = -1$, solid line corresponds to $\omega_0 = 10^{23}$ and dashed line to $\omega_0 = 10^{24}$)

FIG. 7. Models where the sign of $3 + 2\omega$ changes at a finite temperature, $T_*$. On this figure we plot a model with $\Phi(T_*) > 1$ and $\xi$ decreasing monotonically from infinity ($\lambda^2 = 3/2$, $k = -2.5$, solid line corresponds to $\omega_0 = 10^{21}$ and dashed line to $\omega_0 = 10^{23}$). The opposite behavior is also possible with $\Phi(T_*) < 1$ and $\xi$ increasing monotonically from $-\infty$.

FIG. 8. Singular models with $\Phi(0) \to +\infty$ and $\xi$ increasing monotonically from a non-vanishing value of $\xi(0) < 1$ ($\lambda^2 = 3/4$, $k = 0$, solid line corresponds to $\omega_0 = -10^{21}$ and dashed line to $\omega_0 = -10^{23}$)

FIG. 9. Space of theories for a) $\omega_0 > 0, \Phi_0 < 1$ (solid and dashed lines correspond to $\lambda^2/(k+1) = 1$ and 9, respectively), b) $\omega_0 > 0, \Phi_0 > 1$ (the line corresponds to $\lambda/k = 1$), c) $\omega_0 < 0, \Phi_0 < 1$, and d) $\omega_0 < 0, \Phi_0 > 1$.

TABLE 1

Summary of asymptotic solutions

| $3+2\omega(0)$ | $\Phi(0)$ | $3+2\omega \simeq$ | Solution | $\xi(0)$ | |
|---|---|---|---|---|---|
| $>0$ | $+\infty$ | $(3/\lambda^2)(k+x^{-1})\ (k>0)$ | Eq. (34) | $+\infty\ (\lambda/\sqrt{k}<1)$ | S |
| | | | | const. $(\lambda/\sqrt{k}=1)$ | S |
| | | | | $0\ (\lambda/\sqrt{k}>1)$ | S |
| $>0$ | $0$ | $(3/\lambda^2)(k+x^{-1})\ (k>-1)$ | Eq. (37) | $+\infty\ (\lambda/\sqrt{k+1}<1)$ | S |
| | | | | $0\ (1<\lambda/\sqrt{k+1}<3)$ | NS |
| $0$ | $+\infty$ | $(3/\lambda^2)x^{-\epsilon}\ (\epsilon>0)$ | Eq. (39) $(0^+)$ | const. | S |
| | | | Eq. (40) $(0^-)$ | const. $(\lambda>1/2)$ | S |
| $0^+$ | $-\infty$ | $(3/\lambda^2)x^{-\epsilon}\ (\epsilon>0)$ | Eq. (42) | 1 (GR-like) | S |
| $0$ | $0$ | $(3/\lambda^2)(x^{-1}-1)$ | Eq. (45) $(0^+)$ | $-\infty$ | NS |
| | | | Eq. (47) $(0^-)$ | $-\infty\ (\lambda>1/2)$ | NS |
| $<0$ | $<1$ | $-(3/\lambda^2)(x^{-1}-1)$ | Eq. (50) | $-\infty$ | NS |
| $<0$ | $>1$ | $-(3/\lambda^2)(x^{-1}-1)$ | Eq. (40) | $-\infty\ (\lambda\leq 1/2)$ | NS |

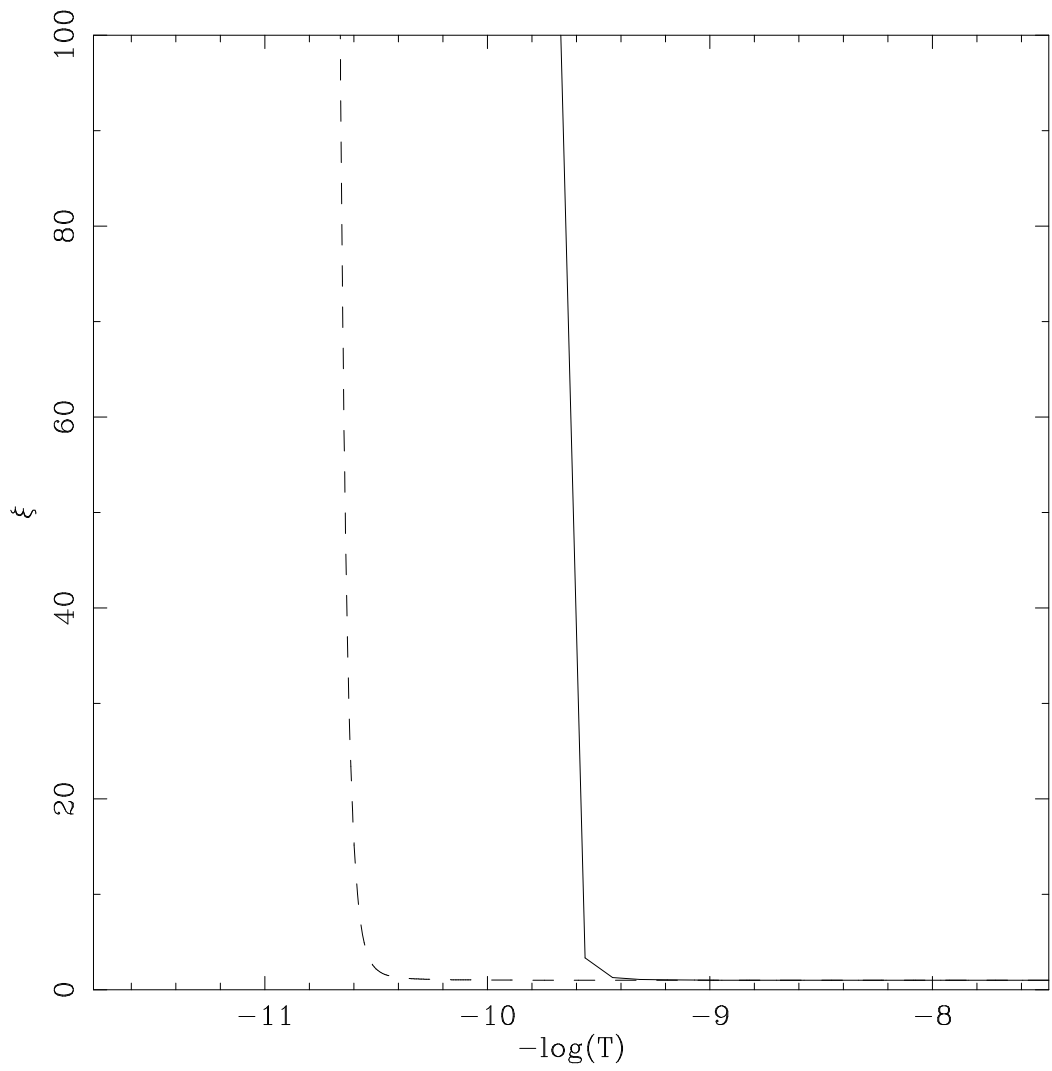

Fig.1

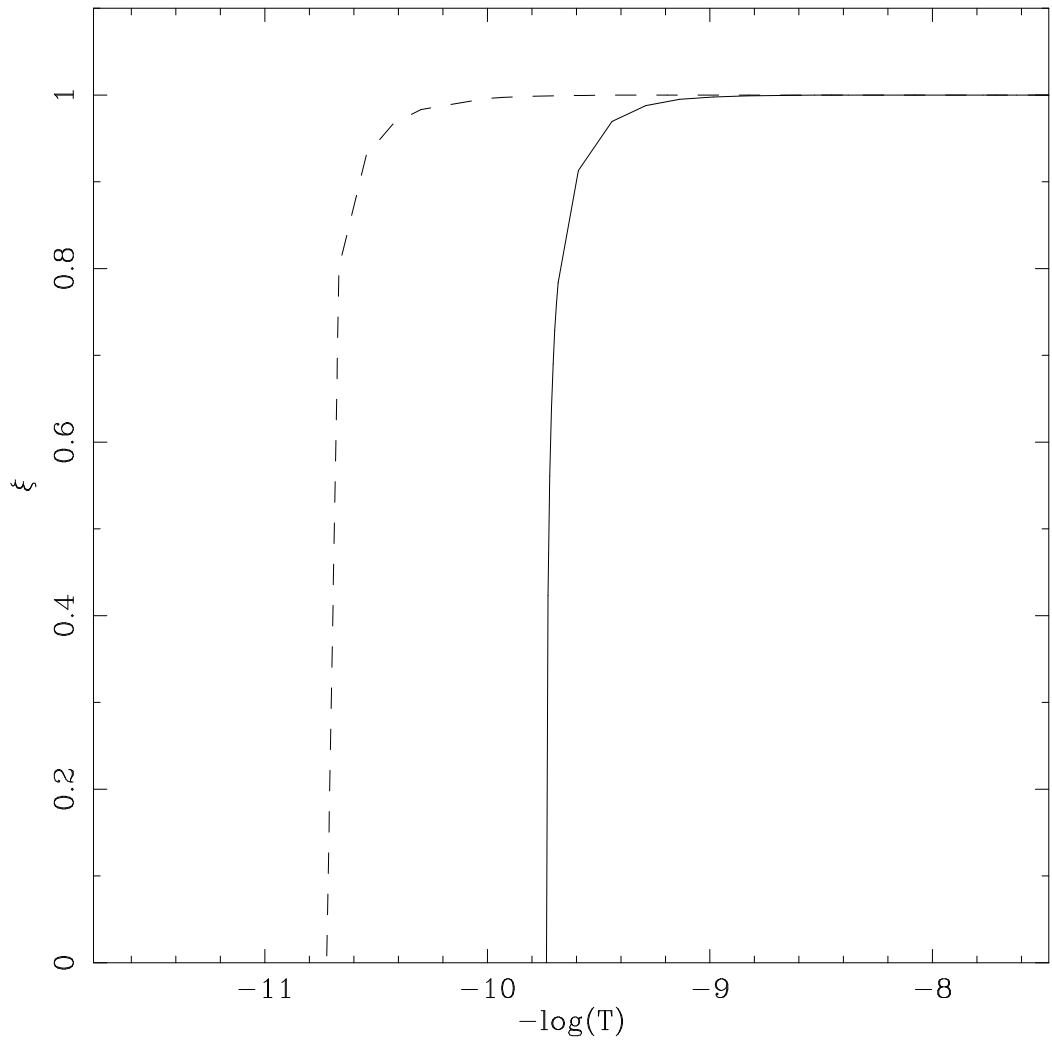

Fig.2

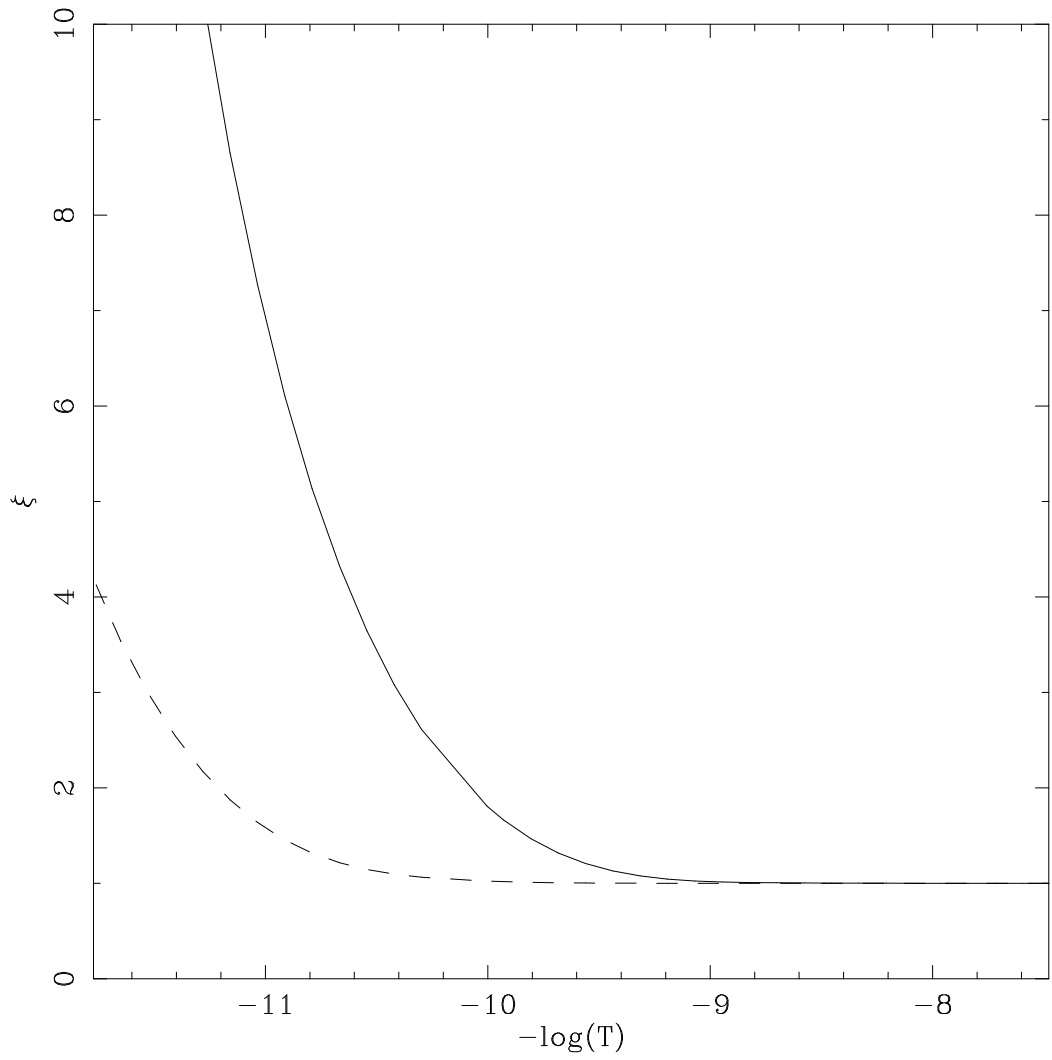

Fig.3

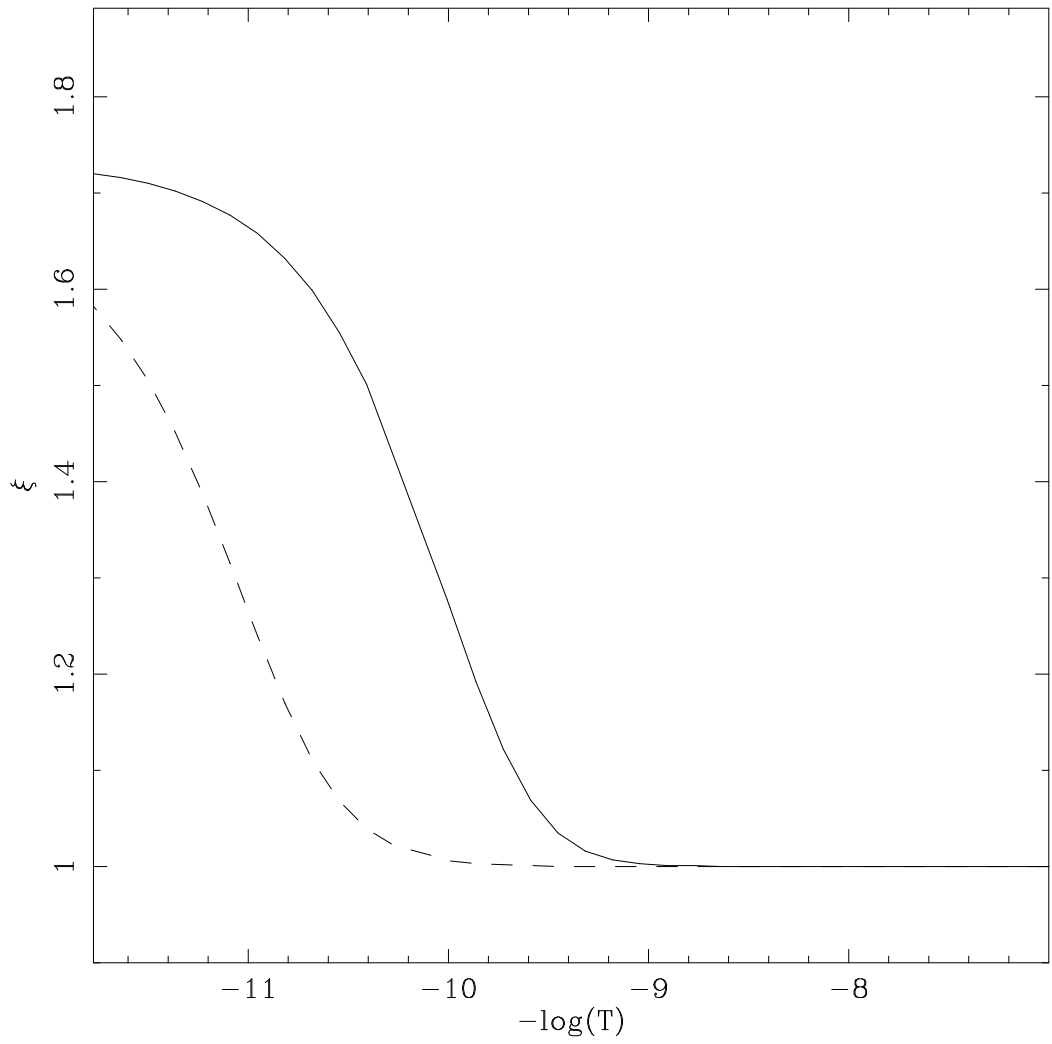

Fig.4

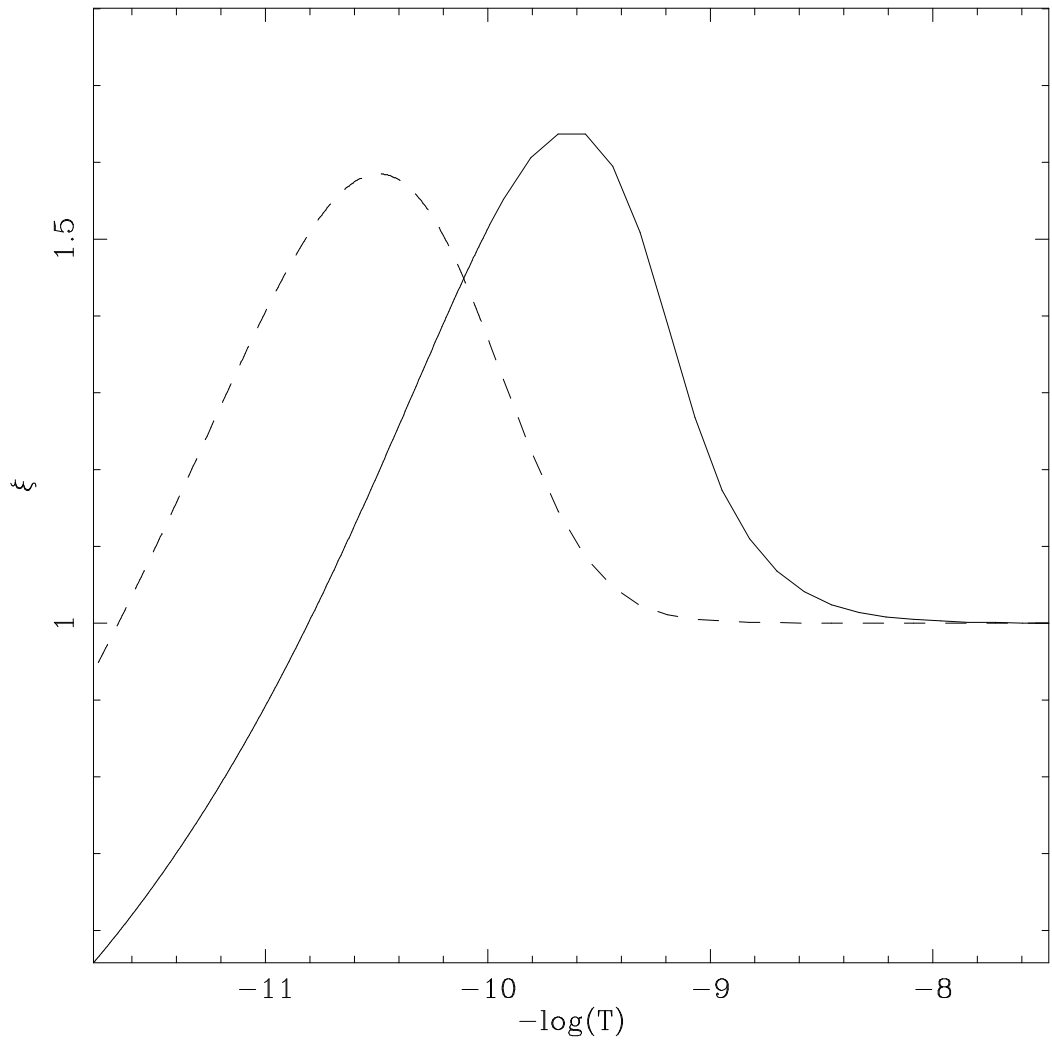

Fig.5

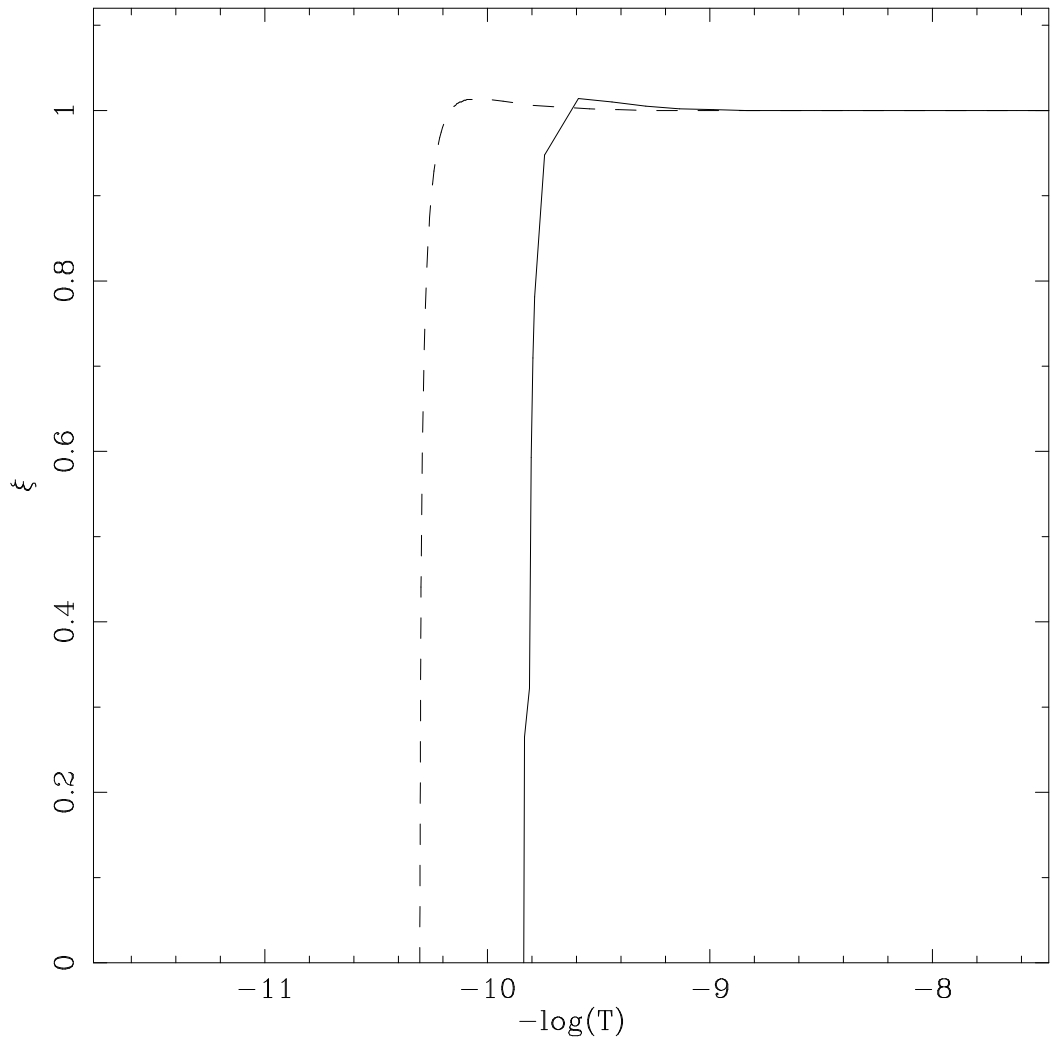

Fig.6

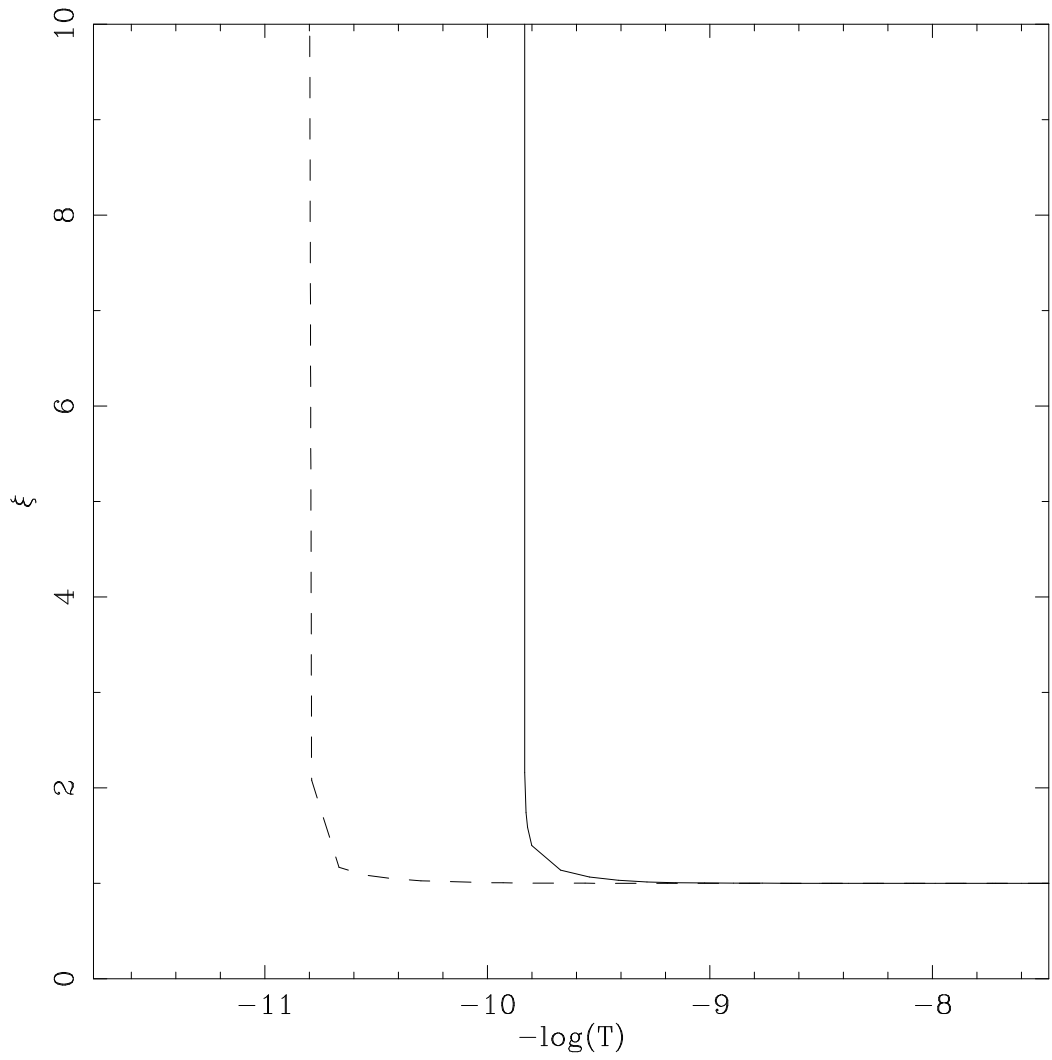

Fig.7

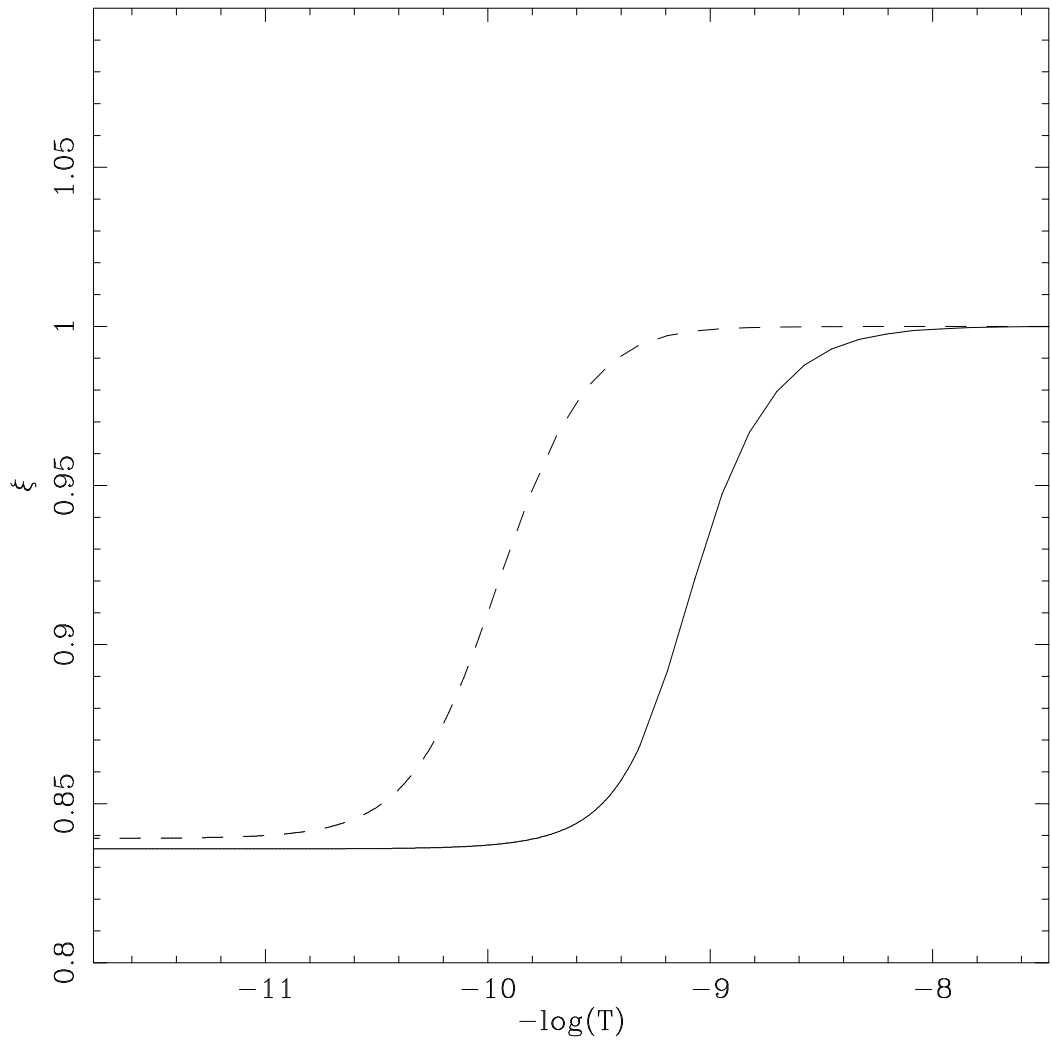

Fig.8

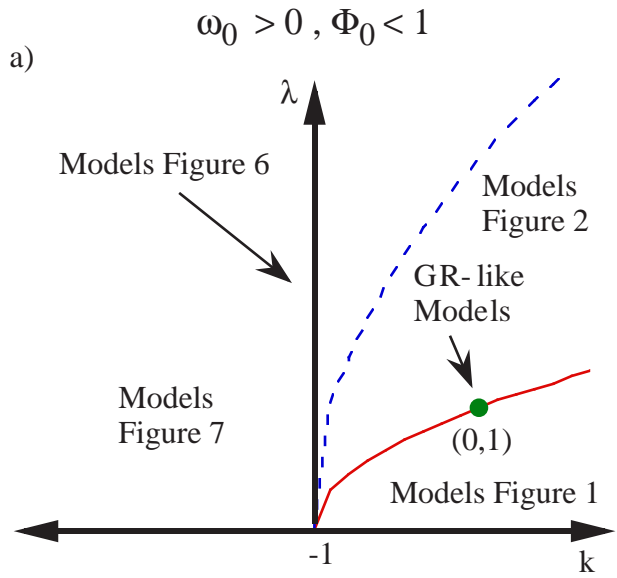
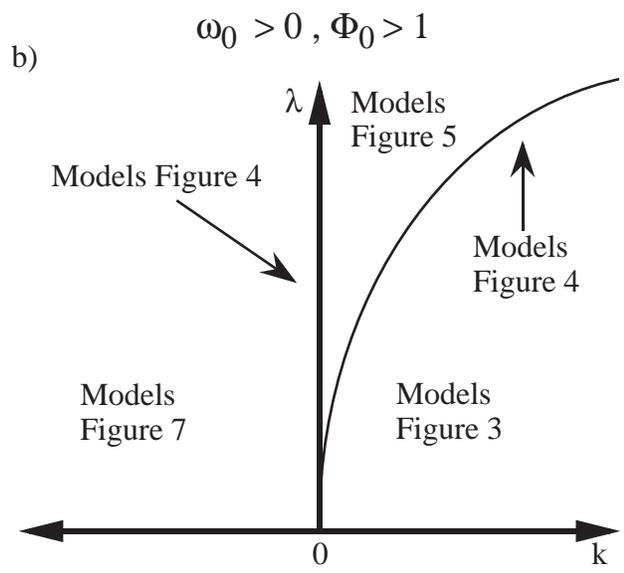
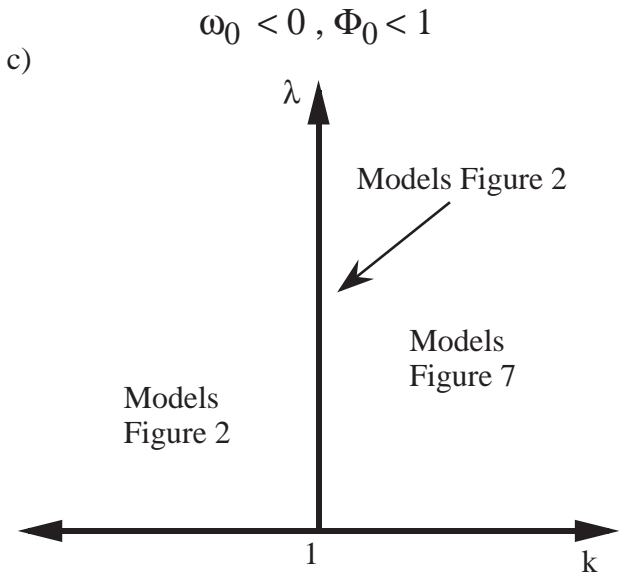
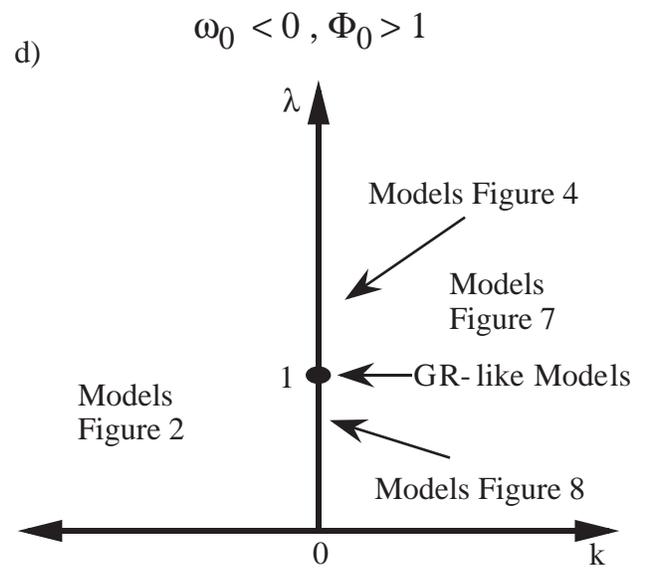

Fig.9